\documentclass[conference]{IEEEtran}
\usepackage[utf8]{inputenc}

\usepackage{tabu}
\usepackage{enumitem}
\usepackage[fleqn]{mathtools}
\usepackage{url}
\usepackage{float}

\title{A Pure HTTP/3 Alternative to MQTT-over-QUIC in Resource-Constrained IoT\vspace{-3mm}}

\author{
  \IEEEauthorblockN{
    Darius Saif\IEEEauthorrefmark{1}, Ashraf Matrawy\IEEEauthorrefmark{2}
  }
  \IEEEauthorblockA{
    Carleton University, Department of Systems and Computer Engineering\IEEEauthorrefmark{1}, School of Information Technology\IEEEauthorrefmark{2}\\
    Email: {Dariussaif\IEEEauthorrefmark{1},Amatrawy\IEEEauthorrefmark{2}}@sce.carleton.ca 
  }
  \vspace{-4mm}
}

\begin{document}
\maketitle

\begin{abstract}
In this paper, we address the issue of scalable, interoperable, and timely dissemination of information in resource-constrained IoT. Scalability is addressed by adopting a publish-subscribe architecture. To address interoperable and timely dissemination, we propose an HTTP/3 (H3) solution that exploits the wide-ranging improvements made over H2. We evaluated our solution by comparing it to a state-of-the-art work: MQTT-over-QUIC. Because QUIC and H3 have undergone standardization in tandem, we hypothesized that H3 would take better advantage of QUIC transport than an MQTT mapping would. Performance, network overhead, and device overhead were investigated for both protocols. Our H3-based solution satisfied our timely dissemination requirement by offering a key performance savings of 1 RoundTrip Time (RTT) for publish messages to arrive at the broker. In IoT networks, with typically high RTT, this savings is significant. On the other hand, we found that MQTT-over-QUIC put marginally less strain over the network.

\end{abstract}


\IEEEpeerreviewmaketitle

\section{Introduction}
To realize an Internet of Things (IoT), many low-cost wireless devices must be deployed in a variety of locations. These sensory devices are often limited in computing resources, storage, and battery. Because of these characteristics, it is imperative that lightweight and energy efficient communication standards be used between sensors themselves, gateways, and fog networking nodes \cite{dizdarevic2019survey}. The heterogeneous nature of IoT makes interoperability ever-more important, giving rise to the Web of Things (WoT) \cite{raggett2015web} for protocols like MQTT and HTTP.


QUIC, an emerging protocol, has become an area of interest in IoT. It is a cross-layer standard that offers integrated security, reduced connection establishment time, and advanced stream multiplexing. It has been found that QUIC can outperform TCP+TLS in networks with higher loss and Round-Trip Time (RTT) \cite{langley2017quic}. Such conditions \textit{are} present in resource-constrained IoT. Researchers \cite{kumar2019implementation,fernandez2020and} have thus mapped MQTT to run over QUIC and evaluated it against MQTT-over-TCP.

QUIC, however, was designed and tuned to carry HTTP traffic, forming the basis of HTTP/3 (H3). Given that, our research objective is to determine whether a pure H3 publish-subscribe (pub-sub) architecture can provide even further gains than MQTT-over-QUIC. We hypothesize that, because of the vertical integration of H3 and QUIC, H3 is better positioned to take full advantage of QUIC transport, instead of MQTT. 

To this end, the contributions of this paper are i.) addressing scalable, interoperable, and timely dissemination of information in resource-constrained IoT networks, and ii.) comparison in three categories between MQTT-over-QUIC and against an H3-enabled pub-sub application of our creation.

\section{Related Work}
Saif \textit{et al.} \cite{saif2021H3} compared web-page performance of H3 draft 27 and H2 in more traditional networks. A suite of metrics was used to shed light on Quality of Experience (QoE). It was shown that H3 was more impervious to network impairment, even though H3's QoE scoring was marginally worse than H2. 

Lars Eggert \cite{eggert2020towards} showed that IoT devices are capable of running QUIC. Two development boards with 32-bit microprocessors were examined running QUIC via  \textit{Quant} and \textit{picoTLS}. Storage space, battery consumption, as well as memory and CPU usage on the boards were monitored while various QUIC transactions (downloads) took place. It was found that QUIC required about 63KB of flash, 16KB of heap memory, 4KB of stack memory, and 0.9J energy per transaction. This was deemed sufficient and that, with further optimizations, QUIC could run on 16-bit processors as well. This work, unlike ours, did not include H3, however.

Kumar and Dezfouli \cite{kumar2019implementation} outfitted Chromium's GQUIC to carry MQTT. Because MQTT and QUIC both run in user-space, they had written inter-process communication APIs and redesigned various data structures to make their solution functional. Raspberry Pi 3 Model B's were used as the subscribers, publishers, and broker. Three network categories were used in their testing: wired, wireless, and long distance. During connection establishment, MQTT-over-QUIC reduced packets exchanged with the broker by up to 56.25\%. To test Head-of-Line Blocking (HoLB), packets were randomly dropped. MQTT (over TCP) was shown to have higher latency in every network configuration. Memory and processor utilization of the broker were also monitored for half-open connections. Although MQTT-over-QUIC used slightly more resources, it relinquished up to 83.24\% and 50.32\% more processor and memory usage respectively compared to MQTT. Lastly, MQTT-over-QUIC was found to be more resilient to when the broker's IP changed mid-transaction.

Fernandez \textit{et al.} \cite{fernandez2020and} had also ported MQTT over to QUIC and compared it against TCP. A pure GO implementation \cite{lucasClem} of QUIC was integrated with Eclipse Paho and VolantMQ. These packages were used as the publisher and broker, respectively. Their implementation was made available open-source. They used NS-3 to emulate different network profiles: WiFi, 4G, and satellite. On each profile, testing of single and multiple MQTT connections were investigated. In the single connection tests, the time delta starting from 1000 packets being published to the last of such messages being pushed to a subscriber was recorded. They found that QUIC outperformed TCP in every configuration, by up to 40\%. Connection establishment time was the focus of the multiple connection testing. After publishing one message, the connection to the publisher and broker was closed. Afterwards, another message was published on a new connection. QUIC proved to perform slightly better than TCP and also exhibited less variation.

Rather than creating specific conversion code and changing data structures to support MQTT over QUIC transport -- which can introduce sub-optimal performance -- our work considers a pure H3 pub-sub implementation. Our implementation was compared against Fernandez \textit{et al.}'s \cite{fernandez2020and} MQTT-over-QUIC.

\section{H3 Publish-Subscribe Architecture}
Previous versions of HTTP were deemed to be inappropriate for IoT due to the nature of its connection management \cite{yokotani2016comparison}. Connection setup times were lengthy and subject to HoLB. Both of these factors were greatly reduced in H3 because of its 0 or 1-RTT establishment times in addition to QUIC's knowledge of streams in the transport layer. Because of these significant gains, we explore the viability of H3 pub-sub.

HTTP was not built for pub-sub, but its APIs are versatile enough to handle it and is also highly interoperable. For the purposes of this paper, an HTTP pub-sub architecture was adapted from Drift \cite{technosophos}. The way Drift handles pub-sub functions over HTTP is shown in Table \ref{tab:h3pubsubsemantics}. In each case below, the client encodes the \textit{topic} name into the URL.

{
\tabulinesep=1mm
\begin{table}[!htb]
  \centering
  \begin{tabu} to 0.48\textwidth {|X[1.5,r]|X[1.1,r]|X[4.7]|}
     \hline
      
      Topic Exists&HEAD&Server returns 200 if it exists, else 404.\\\hline
      
      Create Topic&PUT&Server sanity checks the topic name and creates the topic if checks pass.\\\hline
      
      Delete Topic&DELETE&Server cleans up topic data and unsubscribes all parties from the topic.\\\hline
      
      Publish to Topic&POST&Client encodes topic name in URL and includes data in message body, server stores information and pushes to subscribers.\\\hline
      
      Subscribe to Topic&GET&Client spawns a listener for incoming events, server appends node to the list of nodes subscribed to the topic.\\\hline

 \end{tabu}
  \smallskip
  \caption{H3 Publish-Subscribe Semantic Mappings}
  \label{tab:h3pubsubsemantics}
  \vspace{-5mm}
\end{table}
}

We retooled Drift to run over H3. Drift was chosen due to its few dependencies and open-source code written in GO. Since Fernandez \textit{et al.}'s \cite{fernandez2020and} implementation was also executed with GO, this made for a more fair comparison. In our testing, the underlying QUIC code \cite{lucasClem} was common to H3 pub-sub and MQTT-over-QUIC.

Firstly, we scrapped the entirety of Drift’s \textit{transport} package, because it was based on a GO module of H2 \cite{bradfitz}. We fully replaced and upgraded this functionality with the equivalents from QUIC-GO’s H3 stack package. These changes posed consequences in Drift’s \textit{client} package. This package housed the methods used for mapping pub-sub concepts to HTTP APIs, as outlined in Table \ref{tab:h3pubsubsemantics}. Minor amendments to these methods were necessary in order for us to better accommodate H3 APIs, due to differences in implementation between the H3 and original H2 modules.

Likewise, we also upgraded the \textit{server} package code to run H3. This was taken care of by adding an H3 listener call from QUIC-GO into the \textit{server} package along with a \textit{Cookoo} \cite{cookoo} based HTTP handler. Cookoo, a chain-of-command GO framework, was used for the management of topics and publishers/subscribers on the broker entity.

\section{Network Modelling}

NetEm \cite{hemminger2005network} was leveraged to meet this paper's focus on realistic resource-constrained IoT network conditions. Because Eggert established the validity of running QUIC on IoT devices in \cite{eggert2020towards}, a proof-of-concept approach is used in this paper through the means of emulation. Delay and rate throttling rules were applied to outgoing packets on the publisher and broker network interfaces, as shown in Figure \ref{nbiot}:

\begin{figure}[H]
\centering
\vspace{-1mm}
\includegraphics[width=1.75in]{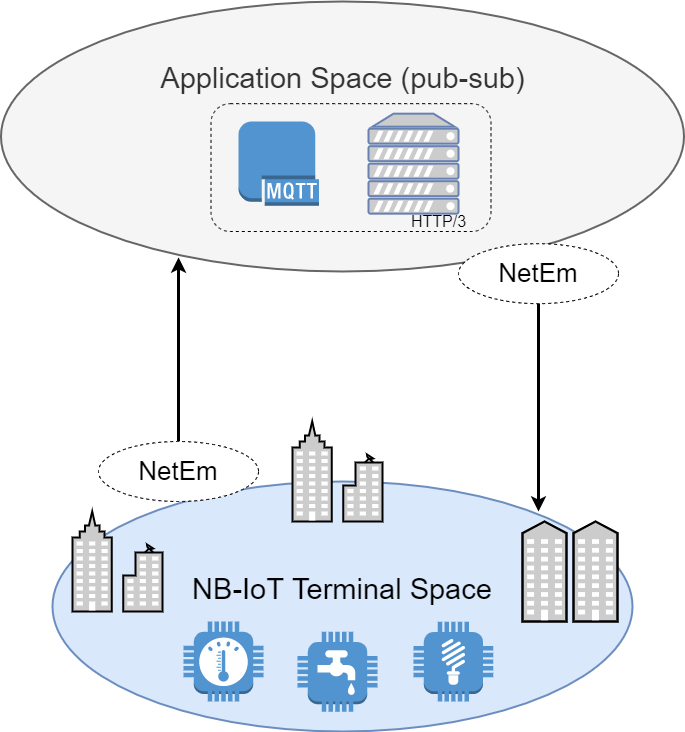}
\vspace{-2mm}
\caption{Emulated Network Topology}
\label{nbiot}
\end{figure}

The 3GPP's Narrowband-IoT (NB-IoT) network parameters were used in this paper. Given NB-IoT's high RTT and modest data rates \cite{zayas20173gpp}, such attributes make it a suitable choice for resource-constrained IoT. The network parameters \cite{zayas20173gpp} of NB-IoT's second generation standard were incorporated into our environment. These parameters are summarized in Table \ref{tab:nbiot}:

{
\tabulinesep=1mm
\begin{table}[!htb]
  \centering
  \begin{tabu} to 0.4\textwidth {|X[3.5,r]|X[4]|}
     \hline
      
      Downlink Rate&127 kbit/s\\\hline
      
      Uplink Rate&159 kbit/s\\\hline
      
      Round-Trip Time&2 seconds\\\hline
      
 \end{tabu}
  \smallskip
  \caption{NB-IoT CAT NB2 Network Parameters}
  \label{tab:nbiot}
  \vspace{-8mm}
\end{table}
}

\section{Experimental System Setup}

Ubuntu 18.04.4 (kernel 4.15.0-88) Virtual Machines (VMs) were deployed in Oracle's VirtualBox for the roles of publishers, subscribers, and the broker. All VMs were allocated 4 processors and 6GB of memory. To allow for VM-to-VM communications, they were configured using a Host-only Ethernet Adapter. The PC hosting the VMs was a 64-bit Windows 10 machine (24GB RAM and Intel i5-836U CPU).

In our setup, default tuning of QUIC-GO \cite{lucasClem} release v0.20.1 powered \textit{both} the H3 and MQTT-over-QUIC implementations. Transport specific factors, like 0-RTT, are expected to equally affect both and are therefore not considered. Thus, we focus on the application layer differences between H3 and MQTT.

The setup adhered to the IETF draft 29 QUIC standard \cite{ietf-quic-http-29} and the \textit{quic\_transport\_parameters} in each protocol's Client Hello were identical. Both also used CUBIC congestion control \cite{ha2008cubic} and path MTU discovery. MQTT-over-QUIC used basic authentication. Dynamic table QPACK was not used for H3; only the static table and encoded string literals were employed. QPACK \cite{ietf-quic-qpack-21} is H3's method of header compression.

NetEm rules provisioned on the broker accounted for the downlink rate and half the RTT value, in accordance to the NB-IoT parameters of Table \ref{tab:nbiot}. Conversely, the uplink rate and half the RTT value from Table \ref{tab:nbiot} were imposed with NetEm on the publisher/subscriber's network interface card.

\section{Results}

\subsection{Performance Indicators}

\subsubsection{Time to First Data Frame} 
\begin{figure}[H]
\centering
\vspace{-3mm}
\includegraphics[width=3.1in]{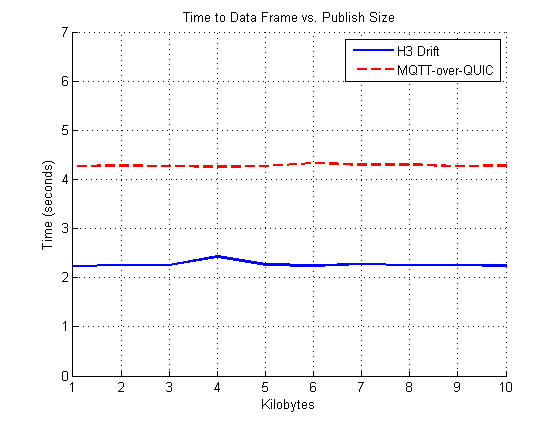}
\vspace{-4mm}
\caption{Time to First Data Frame with Increasing Message Size}
\label{dataStart}
\end{figure}

This was the delta between the publisher's initial Client Hello and its first data frame (H3 DATA FRAME or MQTT PUBLISH packet, respectively). As shown in Figure \ref{dataStart}, this measurement was independent of the size of the message. What was striking, however, was that MQTT-over-QUIC required one additional RTT for the publish message content to be sent -- delaying the message by seconds.

This was due to MQTT's application level connection setup overhead when any form of authentication is used. The first packet a publisher must send to the broker is a CONNECT packet. As per MQTT's specification \textit{"If a Client sets an Authentication Method in the CONNECT, the Client MUST NOT send any packets other than AUTH or DISCONNECT packets until it has received a CONNACK packet"} \cite{mqtt5}. This added an additional RTT which was not necessary in our pure H3 pub-sub implementation.

\subsubsection{Time to Connection Close \& Throughput} Five publisher clients, staggered by one second, each sent one 1KB message to the broker. Packet loss falling within the typical bounds for NB-IoT \cite{tmobile} was applied with a uniform distribution.

\begin{figure}[H]
\centering
\vspace{-3mm}
\includegraphics[width=3.1in]{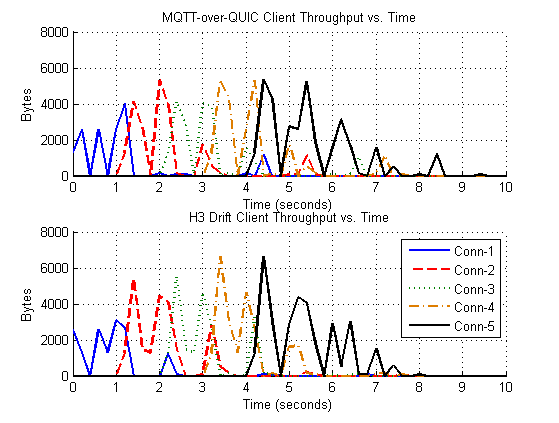}
\vspace{-4mm}
\caption{Interleaved Clients Publishing via MQTT and H3}
\label{throughputInterleaved}
\end{figure}

All of the H3 transactions completed by the 8 second mark, whereas MQTT-over-QUIC lagged roughly 1.4 seconds behind. Also, H3 achieved a 24\% higher peak throughput at the connection level but MQTT-over-QUIC's peak throughput for the aggregate of connections was 2.88\% higher. Figure \ref{throughputInterleaved} shows the connection-level plots for each implementation, with data points aggregated over 200ms intervals.

\subsection{Network Overhead}

\begin{figure}[H]
\centering
\vspace{-3mm}
\includegraphics[width=3.1in]{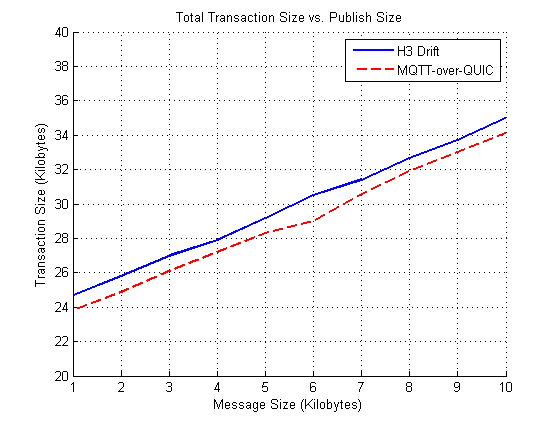}
\vspace{-4mm}
\caption{Transaction Size vs. Increasing Message Size}
\label{captureSize}
\end{figure}

The total amount of bytes and packets exchanged between the broker and publisher was examined for messages of increasing size (1 to 10KB). Figure \ref{captureSize} shows that H3-based transactions required more data exchange than MQTT-over-QUIC -- meaning higher network overhead. Still, the two implementations were quite competitive: the average increase of bytes transmitted was found as 3.23\%. Similarly, the number of packets transmitted for H3 transactions were, on average, 11\% more than MQTT-over-QUIC.

Because of NB-IoT's large RTT, the default tuned QUIC transport was rather aggressive. By the time the broker's response to the publisher's initial establishment request arrived, six additional retry \textit{Client Hello} packets were sent by the publisher. This elicited further server responses, attributing to unnecessary data exchange. Such occurred in every transaction for both implementations (nullifying its effects), which led to the somewhat inflated numbers in Figure \ref{captureSize}.

\subsection{Device Overhead}

Keeping in mind the intended IoT environment, resource consumption was given careful consideration. To this end, the Linux command \textit{ps} was employed in order to give further insights. For the duration of each publish process, \textit{ps} was sampled every 100ms for CPU utilization. 

\begin{figure}[H]
\centering
\vspace{-3mm}
\includegraphics[width=3.1in]{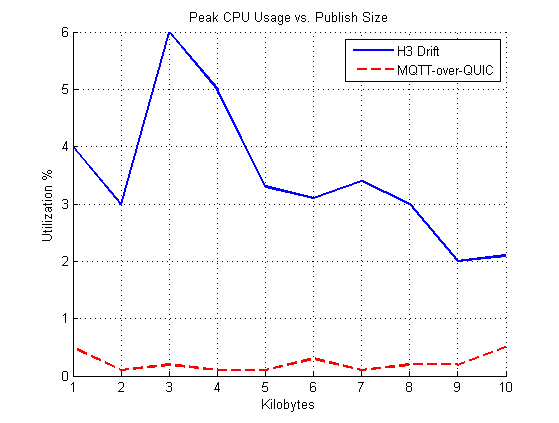}
\vspace{-5mm}
\caption{Peak CPU Usage with Increasing Message Size}
\label{peakCpu}
\end{figure}

The CPU consumption is reported as a percentage -- that is, the ratio of CPU time used to the process' duration. Figure \ref{peakCpu} shows that the CPU usage of H3 was not only unpredictable, but also much higher than that of MQTT-over-QUIC.

\begin{figure}[H]
\centering
\vspace{-2mm}
\includegraphics[width=3.45in]{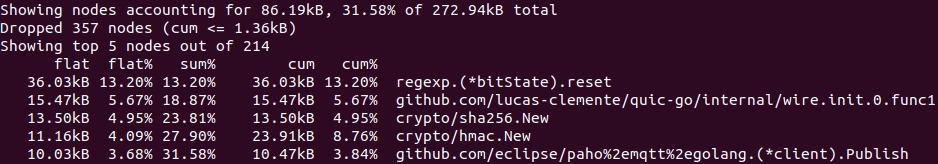}
\vspace{-6mm}
\caption{Top 5 Memory Consumers for MQTT-over-QUIC}
\label{mqtopmem}
\end{figure}

A GO profiler, \textit{pprof}, was used to investigate memory. The total number of bytes allocated in heap memory were investigated for both H3 and MQTT-over-QUIC. The profiler's \textit{MemProfileRate} was driven to 0 in order to record all allocated blocks. Results were collected from 10KB publish messages and the \textit{top} command was used to identify the 5 most memory-hungry functions. These results are shown in Figures \ref{h3topmem} and \ref{mqtopmem}:

\begin{figure}[H]
\centering
\vspace{-2mm}
\includegraphics[width=3.45in]{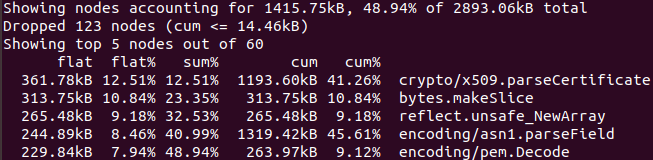}
\vspace{-6mm}
\caption{Top 5 Memory Consumers for H3}
\label{h3topmem}
\end{figure}
\vspace{-1mm}

The \textit{flat} columns represent memory that was allocated, and held, by that particular function whereas \textit{cum} columns are inclusive of the function's children. The \textit{sum\%} column is the summation of the current, and all previous, \textit{flat\%} values.

\vspace{-1mm}

H3 used approximately 10 times the memory as MQTT-over-QUIC. The cause of this disparity is due to H3's TLS configuration. Functions for decoding and parsing the x509 certificate and its fields account for the vast majority of this gap. The x509 certificate overhead also largely attributed to the extra CPU. \textit{pprof} was run in CPU profiling mode for H3 and it uncovered that 37.5\% of H3's time occupying the CPU went towards dealing with the x509 certificate.

\section{Conclusions}
In this paper, an H3 pub-sub alternative to MQTT-over-QUIC for resource-constrained IoT was designed and explored. This implementation retained features akin to MQTT and was quite competitive in terms of network overhead. A savings of 1-RTT during publishing went to H3's favor because of its cross-layer integration with QUIC. The significance of this result is that, in IoT networks, messages can be received (and pushed to subscribers) seconds faster.

H3 was more taxing than MQTT-over-QUIC on the end device -- for resource-constrained IoT, this poses a clear trade-off. It dipped further into the device's CPU and memory. Code profiling with \textit{pprof} in our test environment found this was primarily due to implementation-specific factors.

QUIC/H3 have proven themselves as having strong potential for scalable, interoperable, and timely communication for IoT. In that vein, points of our future work include i.) extending our test environment for more realistic conditions, ii.) non-default stack tuning, and iii.) exploring alternatives to x509 for H3.

\bibliographystyle{ieeetr}
\bibliography{references.bib}

\end{document}